\documentclass{article}

\usepackage{arxiv}
\usepackage[utf8]{inputenc} 
\usepackage[T1]{fontenc}    
\usepackage{hyperref}       
\usepackage{url}            
\usepackage{booktabs}       
\usepackage{amsfonts}       
\usepackage{nicefrac}       
\usepackage{microtype}      
\usepackage{lipsum}
\usepackage{graphicx}
\usepackage{float}
\usepackage[inline]{enumitem}
\usepackage{graphbox}
\usepackage{tikz}
\usepackage{listings}
\usepackage{pythonhighlight}
\usepackage{tabularx}
\usepackage{grffile}
\usepackage{amsmath,mathpazo}
\usepackage{esvect}
\usepackage{algorithm}
\usepackage{adjustbox}
\usepackage{lscape}
\usepackage{wrapfig}
\usepackage{mathtools}
\usepackage{enumitem}
\usepackage{bm}
\usepackage{lscape}
\usepackage{subcaption} 
\usepackage[noend]{algpseudocode}
\makeatletter
\usepackage{geometry}
\usepackage{environ}
\usepackage{color,soul}
\usepackage{setspace}
\usepackage{titlesec}

\title{AuTO: A Framework for Automatic differentiation in Topology Optimization}

\author{
 Aaditya Chandrasekhar \\
  Department of Mechanical Engineering\\
  University of Wisconsin-Madison\\
  Madison, WI \\
  \texttt{achandrasek3@wisc.edu} \\
   \And
 Saketh Sridhara \\
  Department of Mechanical Engineering\\
  University of Wisconsin-Madison\\
  Madison, WI \\
  \texttt{ssridhara@wisc.edu} \\
  \And
 Krishnan Suresh \\
  Department of Mechanical Engineering\\
  University of Wisconsin-Madison\\
  Madison, WI \\
  \texttt{ksuresh@wisc.edu} \\
}

\begin{document}
\maketitle
\begin{abstract}
A critical step in topology optimization (TO) is finding sensitivities. Manual derivation and implementation of the sensitivities can be quite laborious and error-prone, especially for non-trivial objectives, constraints and material models. An alternate approach is to utilize automatic differentiation (AD). While AD has been around for decades, and has also been applied in TO, wider adoption has largely been absent.

In this educational paper, we aim to reintroduce AD for TO, and make it easily accessible through illustrative codes. In particular, we employ JAX, a high-performance Python library for \underline{au}tomatically computing sensitivities from a user defined \underline{TO} problem. The resulting framework, referred to here as AuTO, is illustrated through several examples in compliance minimization, compliant mechanism design and microstructural design. 

\end{abstract}


\section{Introduction}
\label{sec:introduction}
\paragraph{} Fueled by improvements in manufacturing capabilities and computational modeling, the field of topology optimization (TO) has witnessed tremendous growth in recent years. To further accelerate the development of TO, we consider here automating a critical step in TO, namely computing the sensitivities, i.e., computing the derivatives of objectives, constraints, material models, projections and filters, with respect to the design variables, typically the elemental pseudo-densities, in the popular density-based TO.

\paragraph{} Conceptually, there are four different methods for computing sensitivities \cite{baydin2017automatic}: (1) numerical, (2) symbolic, (3) manual, and (4) automatic differentiation. Numerical, i.e., finite difference, based sensitivity computation suffers from truncation and floating-point errors, and is therefore not recommended. Symbolic differentiation using software packages such as SymPy \cite{symPy} or Mathematica ~\cite {Mathematica} is a reasonable choice for simple expressions. However, it is impractical when the quantity of interest involves loops (such as when assembling stiffness matrices), and/or flow-control (if-then-else). The default method today for computing sensitivities is manual. While theoretically  straightforward, the manual process is unfortunately cumbersome and error prone; it is often the bottle-neck in the  development of new TO modules and exploratory studies. In this educational paper, we therefore \emph {illustrate and promote the use of automatic differentiation for computing sensitivities in TO}.

\paragraph{}  Automatic differentiation (AD), is a collection of methods for efficiently and accurately computing derivatives of numeric functions expressed as computer programs \cite{baydin2017automatic}. AD has been around for decades \cite{Rumelhart1986BackProp} and has been exploited in a wide range of problems ranging from  molecular dynamics simulations \cite{schoenholz2019JaxMD} to the design of photonic crystals \cite{minkov2020InversePhotonicCrystalDesign}; see \cite{rall2006perspectivesOnAD} for a critical review.  AD in the context of finite element analysis is reviewed in \cite{ozaki1995higherorderDerivUsingAD} and \cite{van2005reviewSensAnalForTO}.   More recently, AD was demonstrated for for shape optimization  in \cite{paganini2021fireshape} using the Firedrake framework \cite{gangl2020ADforShapeOpt}. AD has also been exploited in TO of turbulent fluid flow systems  \cite{dilgen2018TO_turbulentFlowUsingAD}, \cite{dilgen2018ADforHeatTransfer}.
 
 \paragraph{}  Despite the pioneering research, AD is not widely used in TO.  The objective of this educational paper \emph {is to  accelerate the adoption of AD in TO by providing standalone codes for popular TO problems.} In particular, we employ JAX \cite{jax2018github}, a high-performance Python library for end-to-end AD. Its NumPy \cite{harris2020NumPy} like syntax, low memory footprint and support of just-in-time (JIT) compilation for accelerated code performance makes it an ideal candidate for the task.  We demonstrate the use of AD within the popular density-based TO framework  \cite{bendsoe2013topology}, by replicating existing educational TO codes for compliance minimization \cite{sigmund2001Code99}, compliant mechanism design \cite{Bendsoe2003} and microstructural design \cite{xia2015design}. Critical code snippets are highlighted in this article; the complete codes are available at \href{https://github.com/UW-ERSL/AuTO}{https://github.com/UW-ERSL/AuTO}

\section{ Compliance minimization }
\label{sec:compliance}

\subsection{Problem Formulation }

First we consider compliance minimization as modeled \cite{sigmund2001Code99} in subject to a volume constraint; this TO problem is very popular due to its self-adjoint nature.  In a mesh discretized form, the problem can be posed as:

\begin{subequations}
	\begin{align}
	& \underset{\boldsymbol{\rho}}{\text{minimize}}
	& &J = \boldsymbol{u}^\mathsf{T}\boldsymbol{K}(\boldsymbol{\rho})\boldsymbol{u}\label{eqnObj_compliance}\\
	& \text{subject to}
	& & \boldsymbol{K}(\boldsymbol{\rho})\boldsymbol{u} = \boldsymbol{f}\label{eqn:GoverningEqn_compliance}\\
	& & & \sum_e \rho_e  v_e \leq V^*\label{eqn:volcons_compliance}
	\end{align}
	\label{eq:complianceMinimization}
\end{subequations}
where $\boldsymbol{u}$ is the displacement (in structural problems) or temperature (in thermal problems), $\boldsymbol{K}$ is the stiffness matrix, $\boldsymbol{\rho}$ is the pseudo-density design variables, $\boldsymbol{f}$ is the structural/thermal load and $V^*$ is the volume constraint. To solve this problem, one must define the material model (see below), rely on finite element analysis to solve Equation \ref{eqn:GoverningEqn_compliance}, and use  design update schemes such as  MMA \cite{svanberg1987MMA} or Optimality Criteria \cite{bendsoe1995optimization}.

\paragraph{}A critical ingredient for the design update schemes is the sensitivity, i.e., derivative, of the objective and constraint with respect to the pseudo-density variables. As mentioned earlier, this is typically carried out manually.  For the above self-adjoint problem, the sensitivity of the compliance, for the solid isotropic material with penalization (SIMP) \cite{bendsoe1995optimization} material model, can be easily derived:

\begin{equation}
    \frac{\partial J}{\partial \rho_e} = -\boldsymbol{u}^T \frac{\partial \boldsymbol{K}}{\partial \rho_e}\boldsymbol{u} = -p\rho^{p-1}{u_e}^T K{u_e}
    \label{eq:sensCompliance}
\end{equation}

\emph {However, in this paper,  we will rely on automatic differentiation (AD) framework for sensitivity analysis.}

\subsection{AuTO Framework  }
\paragraph{} The algorithm for solving the above compliance minimization is illustrated in \ref{alg:complianceMinimization}. Code snippets that illustrate the use of AD are discussed below.

\begin{algorithm}[H]
	\caption{Compliance Minimization}		
	\label{alg:complianceMinimization}
	\begin{algorithmic}[1]
		\Procedure{complianceMin}{mesh, material, filter, BC, $V^*$} 
    		\State $i = 0$ \Comment{Iteration index}
    		\State $\rho = V^*$	\Comment{Design variable initialization}
    		\State $\Delta = 1.0$	\Comment{Design change}
    		\While{$\Delta > \epsilon \; \text{and} \; i \leq \text{MaxIter}$}
        		\State $i \gets i+1 $
        		\State $E \gets \rho$ \Comment{Material model} \label{algo:materialModel}
        		\State $\bm{K} \gets E$ \Comment{Compute stiffness matrix and assemble } \label{algo:stiffness}
        		\State $ u \; \text{via} \; Ku=f$ \Comment{Solve with imposed BC} \label{algo:solve}
        		\State $J \gets (K, u)$	\Comment{Objective} \label{algo:compliance}
        		\State $ \frac{\partial \mathcal{J}}{\partial \rho} \gets AD(\rho \rightarrow J)$  \label{algo:AD_objective}	\Comment{Automatic differentiation of objective}
        		\label{algo:constraint}
        		\State $g \gets (\bar{\rho},V^*)$	\Comment{Vol. Constraint}
        		\State $ \frac{\partial g}{\partial \rho} \gets AD(\rho \rightarrow g)$ 	\Comment{Automatic differentiation of constraint}
        		\label{algo:AD_constraint}
        		\State $\phi^i  \gets (J, g , \frac{\partial J}{\partial \rho},\frac{\partial g}{\partial \rho})$ \Comment{MMA Solver \cite{svanberg1987MMA}} \label{algo:callMMA}
        		\State $\Delta =  (||\rho^i - \rho^{i-1}||) $
    		\EndWhile
    		\State \textbf{end while}
		\EndProcedure
		\State \textbf{end procedure}
	\end{algorithmic}
\end{algorithm}

Steps  \ref{algo:materialModel}-\ref{algo:compliance} are captured through the following Python code, where the \textcolor{purple}{@jit} directive refers to the "just-in-compilation", i.e., the compiler translates the Python functions to optimized machine code at run-time, approaching the speeds of C or FORTRAN \cite{lam2015numba}.
\begin{python}
@jit
def computeCompliance(rho):
	E = MaterialModel(rho)
	K = assembleK(E)
	u = solveKuf(K)
	J = jnp.dot(u.T, jnp.dot(K,u))
	return J
\end{python}

SIMP \cite{bendsoe1995optimization} is a typical  material model, and implemented as follows (the \textcolor{purple}{@jit} directive has been removed here to avoid repetition).
\begin{python}
def MaterialModel(rho):
    E = Emin + (Emax-Emin)*rho**penal # SIMP
    return E
\end{python}
The stiffness matrix is assembled in a compact manner as follows.
\begin{python}
def assembleK(E):
    K = jnp.zeros((ndof,ndof))
    sK = D0.flatten()[np.newaxis]*E.T.flatten()
    K = jax.ops.index_add(K, idx, sK)
    return K;
\end{python} 
where $D0 = \int\limits_{\Omega_e}[B]^T[C_0][B] d \Omega_e$ is the element base stiffness matrix \cite{bathe2006finite} with $E = 1$ and prescribed $\nu$; \textit{idx} reflects the global numbering of the element nodes. The underlying linear system is solved using a  direct solver.
\begin{python}
def solveKuf(K): 
    u_free = jax.scipy.linalg.solve(K[free,:][:,free],force[free])
    u = jnp.zeros((ndof))
    u = jax.ops.index_add(u, free,u_free.reshape(-1))
    return u;
\end{python}

Finally, to compute the compliance and its sensitivity in step \ref{algo:AD_objective}, we simply request for the function and its derivative as follows. The JAX environment automatically traces the chain of function calls, and ensures an end-to-end automatic differentiation.
\begin{python}
J,  gradJ = value_and_grad(computeCompliance)(rho)
\end{python}

The global volume constraint (in step \ref{algo:constraint}) is defined as follows, 
\begin{python}
 @jit
 def globalVolumeConstraint(rho):
    vc = jnp.mean(rho)/vf - 1.
    return vc;
\end{python}

As before, the value and its gradient (via AD) can be computed via

\begin{python}
g, gradg = value_and_grad(globalVolumeConstraint)(rho)
\end{python}

As summarized in step \ref{algo:callMMA} of the algorithm, the computed objective, objective gradient, constraint, constraint gradient are then passed to standard optimizers (MMA in our case) \cite{svanberg1987MMA}. The reader is referred to the complete code provided. 

\subsection{Illustrative Examples}
\begin{figure}[!htpb]		
\begin{center}
		\includegraphics[scale=0.6]{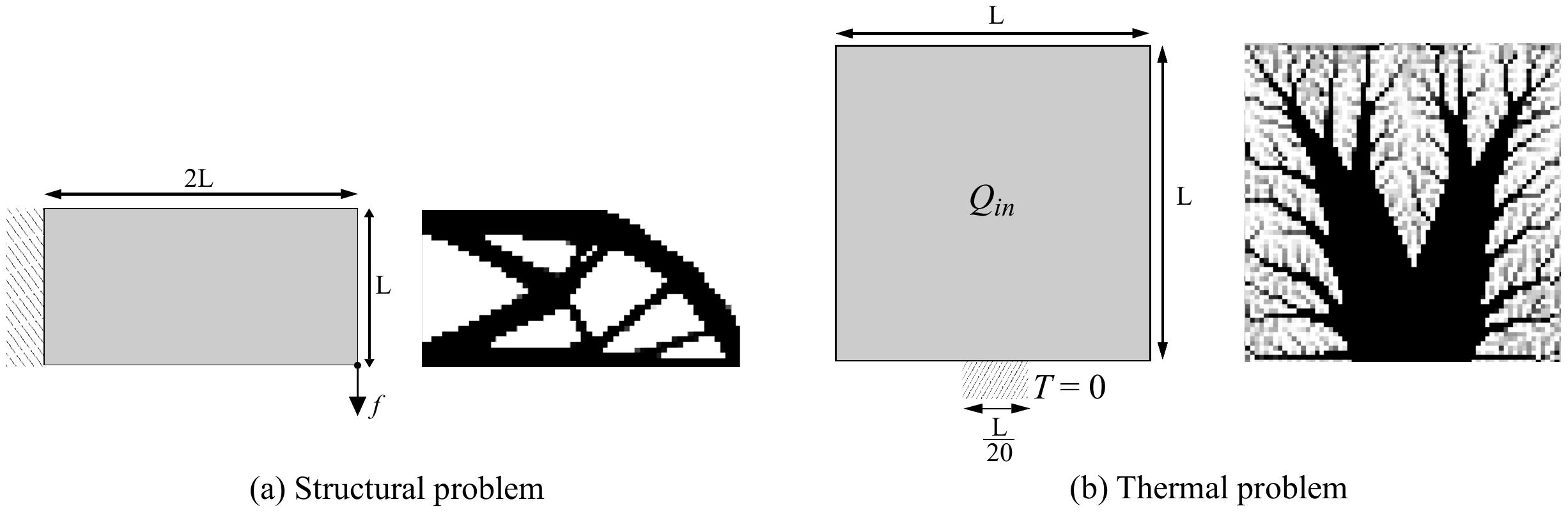}%
		\caption{Compliance minimization examples: (a) Tip loaded cantilever and optimized topology at $V^* = 0.5$ (b) Heat conduction on a square plate and optimized topology at $V^* = 0.5$}
		\label{fig:Compliance_all_BC}
	\end{center}
\end{figure}

\paragraph{} We illustrate the above  AD framework using two popular examples of compliance minimization \cite{bendsoe2013topology}: (a) minimizing structural compliance of a tip-loaded cantilever (see Figure \ref{fig:Compliance_all_BC}a) and (b) minimizing thermal compliance of a square plate under a uniform heat load  (see Figure \ref{fig:Compliance_all_BC}b).  The mesh was chosen to be 60 $\times$ 30  grid for the structural problem, and 60 $\times$ 60 grid for the thermal problem. The target volume fraction in both problems is $V^* = 0.5$. The material properties are $E = 1$, $\nu = 0.3$, $k = 1$, and  MMA was used as the design update scheme, with default parameters. The computed designs illustrated in \ref{fig:Compliance_all_BC}a and \ref{fig:Compliance_all_BC}b matches those in the literature \cite{bendsoe2013topology}.

\begin{figure}[H]
	\begin{center}
		\includegraphics[scale=0.5,trim={100 90 30 80},clip]{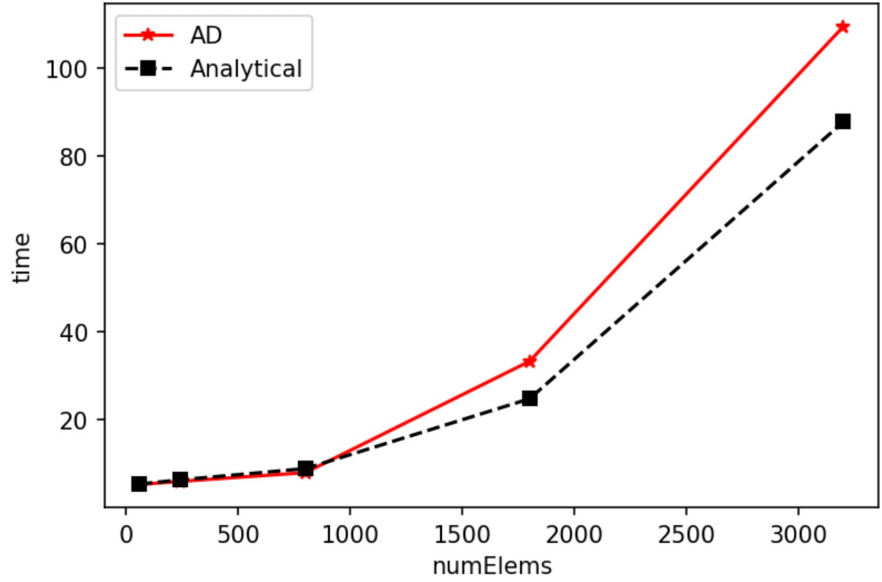}%
		\caption{Computational cost for compliance minimization using AD and analytical implementation.}
		\label{fig:timing_compliance}
	\end{center}
\end{figure}

The totla time taken for optimization using analytical derivatives and AD are compared in Figure \ref{fig:timing_compliance}.  We observe that  AD is marginally more expensive, but we will observe later that this is not always the case.

\subsection{Advantages of AD}
\paragraph{} While SIMP is a popular material model, other models have been proposed \cite{dzierzanowski2012comparisonSIMPvRAMP}. The advantage of AD is that one can easily replace  SIMP with, for example, RAMP ~\cite{stolpe2001RAMP}, by simply changing the material model.
\begin{python}
def MaterialModel(rho):
    E = Emax*rho/(1.+S*(1.-rho)) # RAMP
    return Y
\end{python}
All downstream sensitivity computations are handled automatically.

\paragraph{} Often additional filters and projections are used in TO. For instance, they can be used to remove checkerboard patterns \cite{Sigmund1998NumericalInstabTO}, impose minimum length scale \cite{guest2004MinLengthScaleProjection}, limit gray elements \cite{Wu2017}, etc. The filters apart from being  complex in their own right, they are often used in tandem. For instance, in \cite{Wu2017ShellInfill}, eight such schemes were compounded to obtain shell-infill type structures. This results in highly complicated sensitivity expressions that can be laborious to derive. However, using an AD framework, the user simply needs to include the desired projections in the pipeline and the sensitivity is taken care of. 
For instance, we can introduce the following   filter to reduce grayness in design,  just before computing the material model. 
\begin{python}
def projectionFilter(rho):
    if(projection['isOn']):
        nmr = np.tanh(c0*beta) + jnp.tanh(beta*(rho-c0))
        dnmr = np.tanh(c0*beta) + np.tanh(beta*(1-c0))
        rho_tilde = nmr/dnmr
        return rho_tilde
    else:
        return rho
\end{python}

\begin{figure}[h!]
	\begin{center}
		\includegraphics[scale=0.25,trim={10 90 0 70},clip]{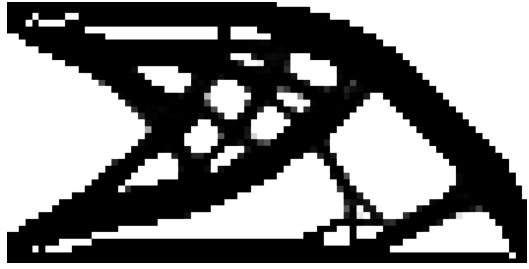}%
		\caption{Tip cantilever beam  with length scale control.  }
		\label{fig:tipCantilever_50vf_maxLS}
	\end{center}
\end{figure}

\paragraph{} Finally,  manufacturing constraints  \cite{vatanabe2016ManufCons}, \cite{liu2018current} are often imposed in TO; these include limiting overhang of structures \cite{Qian2017}, connectivity \cite{li2016structuralConnectivityConstraint}, material utilization \cite{Sanders2018Multimaterial}, and length scale control \cite{Guest2009MaxLengthScale}. Such constraints are easy to impose within the AD framework. For example, the local volume constraint proposed in \cite{Guest2009MaxLengthScale} may be implemented as follows. 
\begin{python}
def maxLengthScaleConstraint(rho):
    v = jnp.matmul(L, (1.01-rho)**n); # L averaged prior 
    cons = 1 - jnp.power(jnp.sum(v**p),1./p)/vstar;
    return cons;
\end{python}
As before, one can calculate the value and gradient of the constraint via
\begin{python}
vc, gradvc = value_and_grad(maxLengthScaleConstraint)(rho)
\end{python}
The computed constraint and gradient can then be passed on to MMA. To illustrate, for the tip cantilever problem  in Figure \ref{fig:Compliance_all_BC}(a), with the additional max length scale radius of $r = 30$, and maximum void volume at $0.75 \pi r^2$, the resulting topology is illustrated in Figure \ref{fig:tipCantilever_50vf_maxLS}.

\section{Compliant Mechanism Design}
\label{sec:compliantMechanism}
We next illustrate the AuTO framework using compliant mechanisms (CMs) \cite{howell2013compliant}; see \cite{zhu2020design} for a comprehensive review on TO for CMs.

\subsection{Problem Formulation}
\label{sec:compmech_probFormulation}
\begin{figure}[h!]
	\begin{center}
		\includegraphics[scale = 0.8]{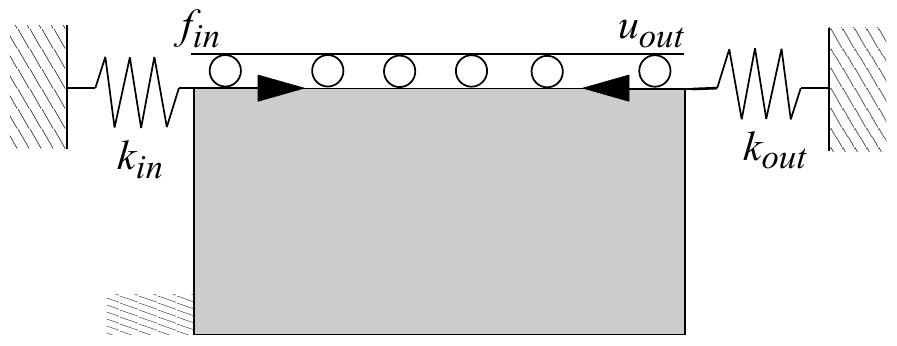}
		\caption{The displacement inverter compliant mechanism.}
		\label{fig:Compliant_Mech_BC}
	\end{center}
\end{figure}

Consider the displacement inverter considered in the 104-line educational MATLAB code \cite{bendsoe2013topology}. The objective is to maximize the output displacement $u_{out}$ at the point of interest when a force $f_{in}$ is applied, as illustrated in Figure \ref{fig:Compliant_Mech_BC}. The spring constants are specified by the user to control the behavior of the CM.
 \\
This TO problem can be written as:
\begin{subequations}
	\begin{align}
	& \underset{\boldsymbol{\rho}}{\text{maximize}}
	& &\boldsymbol{u}_{out}\label{eqnObj_compliantMech}\\
	& \text{subject to}
	& & \boldsymbol{K}(\boldsymbol{\rho})\boldsymbol{u} = \boldsymbol{f_{in}}\label{eqn:GoverningEqn_compliantMech}\\
	& & & \sum_e \rho_e  v_e \leq V^*\label{eqn:volcons_compliantMech}
	\end{align}
\end{subequations}
The standard implementation entails computing  the elemental sensitivity $\frac{\partial u_{out}}{\partial \rho_e}$ given by:
\begin{equation}
    \frac{\partial u_{out}}{\partial \rho_e} = \boldsymbol{\lambda}^T \frac{\partial \boldsymbol{K} }{\partial \rho_e}\boldsymbol{u} = p\rho^{p-1}{\lambda_e}^T K{u_e}
\end{equation}
where $\boldsymbol{\lambda}$ is the solution to the adjoint load problem $\boldsymbol{K\lambda} = -\boldsymbol{l}$. $\boldsymbol{l}$ is a vector with the value 1 at the degree of freedom corresponding to the output point, and with zeros elsewhere. Observe that, in the manual method,  two sets of analysis (one to compute $\boldsymbol{u}$, and the other to compute $\boldsymbol{\lambda}$) are required per iteration for evaluating sensitivities.

\subsection{AuTO Framework}
The implementation in AuTO for CM design is similar to the compliance minimization problem, with two minor changes: (a) the stiffness matrix assembly includes the spring constants, and (b) the objective is the displacement at the output node.
The relevant code snippets are provided below.
\begin{python}
def assembleKWithSprings(E):
    K = jnp.zeros((ndof,ndof))
    sK = D0.flatten()[np.newaxis]*E.T.flatten()
    K = jax.ops.index_add(K, idx, sK)
    # springs at input and output nodes
    K = jax.ops.index_add(K, jax.ops.index[nodeIn, nodeIn], kspringIn)
    K = jax.ops.index_add(K, jax.ops.index[nodeOut, nodeOut], kspringOut)
    return K;
\end{python}
\begin{python}
def CompliantMechanism(rho):
	E = MaterialModel(rho)
	K = assembleKWithSprings(E)
	u = solveKuf(K)
	return u[bc['nodeOut']]
\end{python}

To compute the objective and its gradient, we rely on JAX as follows.
\begin{python}
J,  gradJ = value_and_grad(CompliantMechanism)(rho)
\end{python}

The design update using MMA is as per Section \ref{sec:compliance}. Using the problem specification in \cite{Bendsoe2003}, the resulting  topology for the inverter is illustrated in Figure \ref{fig:output_compliant_mech_all}a; this is in agreement with the result in  \cite{Bendsoe2003}.

\subsection{Advantages of AD}

For the design of CMs using TO, a key advantage of AD stems from the following observation \cite{zhu2020design} "\textit{no universally accepted objective formulation exists}". For example, consider two additional objectives:
    \begin{enumerate}
        \item $\min: -\omega MSE + (1 - \omega)SE$ \cite{nishiwaki1998topology}
        \item $\min: -MSE/SE$ \cite{saxena2000optimal}
    \end{enumerate}
where $MSE = \boldsymbol{v}^T\boldsymbol{K}\boldsymbol{u}$ is the mutual strain energy, which describes the flexibility of the designed mechanism and $SE = \boldsymbol{u}^T\boldsymbol{K}\boldsymbol{u}$ is the strain energy, $\boldsymbol{v}$ is the output displacement when a unit dummy load applied at the degree of freedom corresponding to the output point. 

\begin{figure}[H]
	\begin{center}
		\includegraphics[scale=0.6]{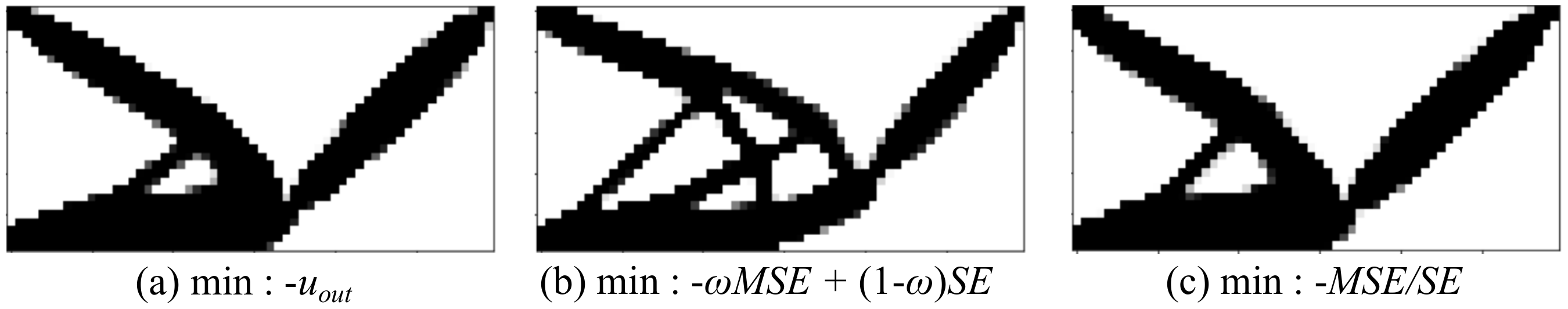}%
		\caption{Displacement inverter design using three formulations at $V^* = 0.35$}
		\label{fig:output_compliant_mech_all}
	\end{center}
\end{figure}

\begin{figure}[H]
	\begin{center}
		\includegraphics[scale=0.5,trim={100 90 30 80},clip]{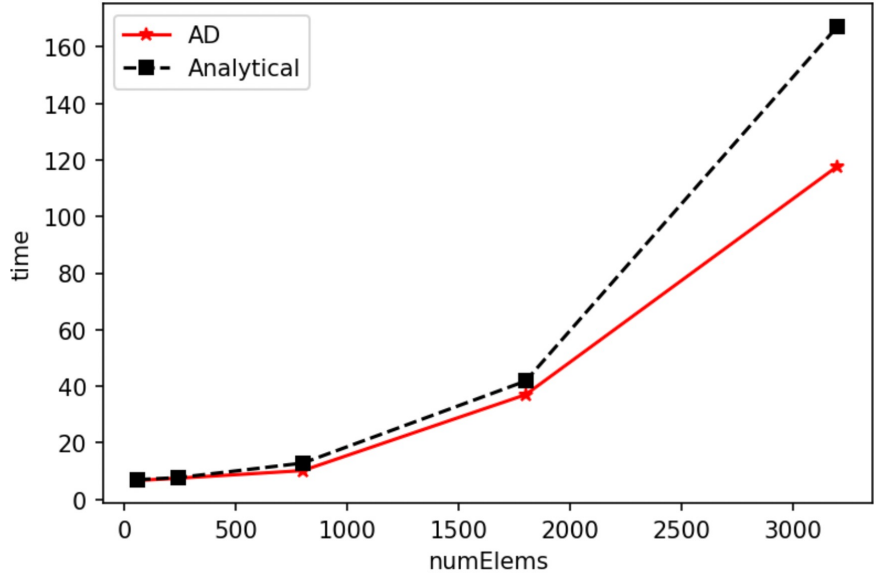}%
		\caption{Computational cost of optimization using AD vs analytical implementation. }
		\label{fig:timing_call_CM}
	\end{center}
\end{figure}

In the AuTO framework, one can easily explore various objectives as follows.
\begin{python}
def CompliantMechanism(rho):
	E = MaterialModel(rho)
	K = assembleKWithSprings(E)
	u = solveKuf(K)
	v = solve_dummy(K)
	MSE = jnp.dot(v.T, jnp.dot(K,u))
	SE = jnp.dot(u.T, jnp.dot(K,u));
	J = -MSE/SE # or 
	# w = 0.9
	# J = -w*MSE + (1 - w)*SE 
	return J
\end{python}
The topologies obtained with the two additional objectives  are illustrated in Figure \ref{fig:output_compliant_mech_all}b and Figure \ref{fig:output_compliant_mech_all}c.
For the objective of maximizing output displacement, the computational costs using analytical and AD methods  are illustrated in Figure \ref{fig:timing_call_CM}. Observe that the the analytical method is more expensive since one must solve an adjoint problem explicitly. On the other hand, JAX internally optimizes the code for computing sensitivities via AD.

\section{Design of Materials}
\label{sec:microstructuralDesign}
In this section, we replicate the educational article \cite{xia2015design} for the design of microstructures using AuTO. In particular, we consider (a) maximizing bulk modulus (b) maximizing shear modulus, and (c) designing microstructures   with negative Poisson's ratio.

\subsection{Problem setup}
The mathematical formulation is as follows \cite{xia2015design}:

\begin{subequations}
	\begin{align}
	& \underset{\boldsymbol{\rho}}{\text{minimize}}
	& &c( E_{ijkl}^H( \boldsymbol{\rho}))\label{eqnObj_microstr}\\
	& \text{subject to}
	& & \boldsymbol{K}(\boldsymbol{\rho})\boldsymbol{U}^{A(kl)} = \boldsymbol{F}^{(kl)}, k,l= 1,2,\ldots,d\label{eqn:GoverningEqn_microstr}\\
	& & & \sum_e \rho_e v_e \leq V^*\label{eqn:volcons_microstr}
	\end{align}
\end{subequations}
where the objective $c(E_{ijkl}^H)$ represents the material property we intend to minimize,  $\boldsymbol{K}$ is the  stiffness matrix, $\boldsymbol{U}^{A(kl)}$ and $\boldsymbol{F}^{(kl)}$ are the displacement vector and the external force vector for  test case $(kl)$ respectively. The different test cases correspond to the unit strain tests along different directions, where $d$ is the spatial dimension.

\paragraph{}In 2D, maximization of bulk modulus corresponds to:
\begin{equation}\label{eq:bulk}
    c = -(E_{1111}+E_{1122}+E_{2211}+E_{2222})
\end{equation}
and maximization of shear modulus corresponds to:
\begin{equation}\label{eq:shear}
    c = -E_{1212}
\end{equation}
Finally, for the design of materials with negative Poisson's ratio, the following was proposed  \cite{xia2015design}:
\begin{equation}\label{eq:poisson}
    c = -E_{1122} - \beta^l(E_{1111}+E_{2222})
\end{equation}
where $\beta \in (0,1)$ is a user-defined fixed parameter and $l$ is the design iteration number. Observe that, in the manual method, computing the sensitivity requires solving for the adjoint \cite{xia2015design}, 

\subsection{Implementation on AuTO}

\begin{figure}[h!]
	\begin{center}
		\includegraphics[scale=0.5,trim={10 240 20 80},clip]{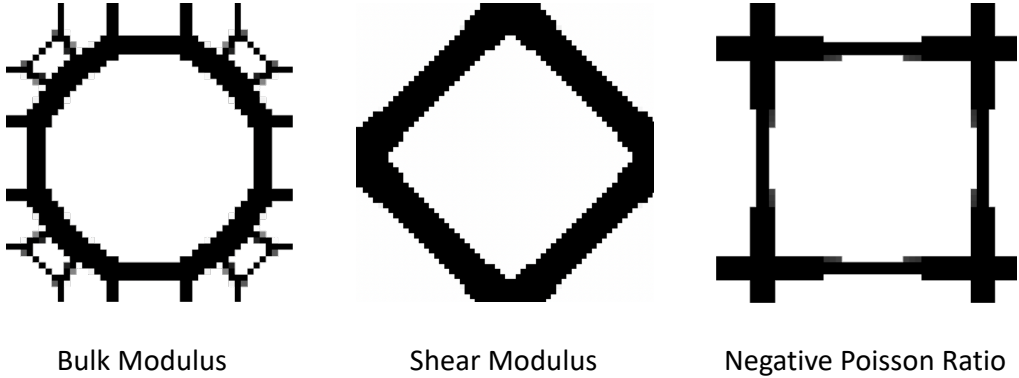}%
		\caption{Maximization of bulk modulus, shear modulus and design of material with negative Poisson's ratio, with $v_f^* = 0.25$.}
		\label{fig:method_fiberOptimization}
	\end{center}
\end{figure}

In AuTO, the bulk modulus objective, for example, can be captured as follows: 

\begin{python}
def MicrosructuralDesign(rho):
    E = MaterialModel(rho)
    K = assembleK(E)
    Kr, F = computeSubMatrices(K)
    U = performFE(Kr, F)
    EMatrix = homogenizedMatrix(U, rho)
    bulkModulus = -EMatrix['0_0']-EMatrix['0_1']-EMatrix['1_1']-EMatrix['1_0']
    return bulkModulus
\end{python}
Other objectives can be similarly captured. Figure \ref{fig:method_fiberOptimization} illustrates three different microstuctures for the three different objectives, for a volume fraction of 0.25. 

\section{Conclusion}
\label{sec:conclusion}

\paragraph{}In this paper, we demonstrated the simplicity and benefits of AD in TO. Possible extensions include multi-physics \cite{alexandersen2020review, semmler2018material, deng2017TO_thermoelasticBuckling} and non-linear problems \cite{wang2014design,clausen2015topology}. In the current implementation, 
direct solvers were employed. For large scale problems, sparse pre-conditioned iterative solvers \cite{andersen2013cvxopt}, \cite{Yadav2013AssemblyFree} will be critical (but not fully supported by JAX). One of the advantages of the manual approach (that is lost in AD) is that the expressions can provide key insights to the problem.

\section*{Replication of results}
The Python code pertinent to this paper is available at \href{https://github.com/UW-ERSL/AuTO}{https://github.com/UW-ERSL/AuTO}.

\section*{Acknowledgments}
The authors would like to thank the support of National Science Foundation through grant CMMI 1561899.

\section*{Compliance with ethical standards}
The authors declare that they have no conflict of interest.

\bibliographystyle{unsrt}  
\bibliography{AuTO,ERSLReferences,TOuNN_biblio} 
\end{document}